# POSSIBLE MATHEMATICAL FORMULATION OF LIQUIDITY PREFERENCE THEORY

ALEXANDER MOROZOVSKY

Bridge, 57/58 Floors, 2 World Trade Center,
New York, NY 10048

ABSTRACT

New mathematical formulation of liquidity preference theory is suggested. On the base of comparison between suggested model and real prices paradoxical conclusion could be derived. The whole yield curve could be described only on the base of liquidity preference theory without any consideration of risk-neutral models.

One of the existing problems in the theory of mathematical finance is the calculation of term structure of interest rates. The term structure of interest rates is determined by the dependence of the yield of the discount instruments from maturity.

There are many different models explaining the relationship between long- and short-term interest rates [1]. One of the most popular is the liquidity preference theory. This theory states that forward rates should be higher than future spot rates. This means that long-term buyers of bonds require additional return. Let's rephrase our statements in such a way so that it would be similar to concepts from portfolio theory (return connected with risk). So, we could assume that holders of long-term securities receive additional return compared with holders of short-term securities because of the additional risk associated with long-term securities. Because of the resemblance of this statement to the portfolio theory, we could try to apply powerful mechanisms of portfolio analysis in order to calculate the yield curve.

We could formulate again that the purpose of this paper is the mathematical formulation of liquidity preference theory on the base of similarity of concepts, underlying this theory with portfolio analysis.

In order to write an equation for yield calculation we will apply the following important theoretical concepts:

1. Markowitz portfolio theory for connection between risk and return [2].
2. Value-at-risk concept for measuring risk [3].
3. Creation of riskless portfolio as a tool for obtaining riskless return [1].

The simplest reformulation of liquidity preference theory is: additional risk requires additional return, or

$$\text{Additional Return} = \alpha * \text{Additional Risk} \quad (1)$$

Now we will try to elaborate our statement on the base of concepts from the modern theory of finance, mentioned above. The remaining part article consists of the following parts:

1. How we are going to calculate additional return.
2. What is the reason for additional risk and how are we going to calculate it.
3. Final form of the basic equation and its solution.
4. Comparison of empirical prices with obtained equation.
5. Correspondence between suggested approach and portfolio theory.
6. Conclusion.

### 1. How we are going to calculate additional return.

We will define additional return as the difference between the market price of bond($P_r$) and the price of the bond, calculated from riskless approach(E):

$$\text{Additional return} = E - P_r, \quad (2)$$

(E - $P_r$, not $P_r$ - E). The following order becomes clearer if we write down an additional return as difference between the return for a real bond and return in the risk-neutral world:

$$\text{Additional return} = \text{Return for bond} - \text{Return for bond in risk-neutral world}. \quad (3)$$

Consider $I$ as the value of all cash flows, connected with the bond at the time of maturity, we could define the terms in (3) as:

$$\text{Return for bond} = I - P_r,$$
$$\quad (4)$$
$$\text{Return for bond in risk-neutral world} = I - E,$$

Because of (3) and (4) additional return could be written as:

$$\text{Additional return} = (I - P_r) - (I - E), \quad (5)$$

and we could immediately see that (2) and (5) are the same equations. The calculation of additional risk, however, is much more difficult.

## 2. What is the reason for additional risk and how we are going to calculate it.

It's possible to suggest many reasons for risks, existing even if we use the usual risk - neutral framework. First of all, it could be transaction costs (or more general - market could be incomplete). Then, it could be imprecision of used risk-neutral models or the existence of many of them. And, of course, it could be the usual arguments for liquidity preference theory. In this explanation, additional risk depends not only on volatility $\sigma$, but also on time to maturity. We need some additional quantitative concept for measuring this additional risk. All of the following reasons could lead to deviation from the risk-neutral approach and to the existence of additional risk.

In order to estimate an additional risk we will use Value-at-Risk concept developed at J. P. Morgan ([3]). According to this concept, Value-at-Risk (VaR) is equal to the difference between average expected price at time T (time of maturity) and price of instrument that differentiate probability space in a special way(such that, the integral probability to be below the average expected price of this instrument would be n%, where n=1,2,3). Now, we will specify equation for interest-rate security

$$dF = \mu_F F dt + \sigma_F F dz \tag{6}$$

where F - the value of security, z - Wiener's process, and $\mu_F$, $\sigma_F$ are generally speaking - functions from F and T.

Now, we will apply usual formula for VaR for security, described by (6) and measure additional risk as quantity proportional to VaR:

$$AdditionalRisk \approx P_r * e^{\mu^e \Delta t}(1 - e^{-\beta \sigma^e \sqrt{\Delta t}}), \tag{7}$$

Where $\mu^e$, $\sigma^e$ - some functional from $\mu_F$ and $\sigma_F$ correspondingly. It's possible to consider different approaches for $\mu^e$ and $\sigma^e$ calculating. We will consider formula (7) when instead of $\mu^e$ and $\sigma^e$ we will use averages in time:

$$\mu^e = \frac{1}{\Delta t} \int \mu_F \, dt \tag{7-1},$$

and

$$\sigma^{e^2} = \frac{1}{\Delta t} \int \sigma_F^2 dt \qquad (7\text{-}2)$$

where (7-1) and (7-2) are the simplest average characteristics for $\mu_F$ and $\sigma_F$.

We should outline that instead of using VaR ideas for Additional Risk (7) it is possible to suggest different definitions for Additional Risk.

The simplest forms of dependence of $\sigma^e$ from $\Delta t$ would be:

$$\sigma^e \approx \sigma_0 \qquad (8)$$

and

$$\sigma^e \approx \sigma_1 \Delta t \qquad (9)$$

when the volatility $\sigma_F$ (8) or its derivative (9) is constant.

## 3. Final form of basic equation (connection between additional risk and additional return and its solution).

There is, however, one small problem, that needs to be solved in order to write final version of equation (1) (connection between additional risk and additional return). The problem is the following: payoff for additional risk happened immediately at time t = 0(additional return), but this risk is calculated at time of maturity( at t = T, Fig.1):

**Fig.1 Time diagram for additional risk and additional return.**

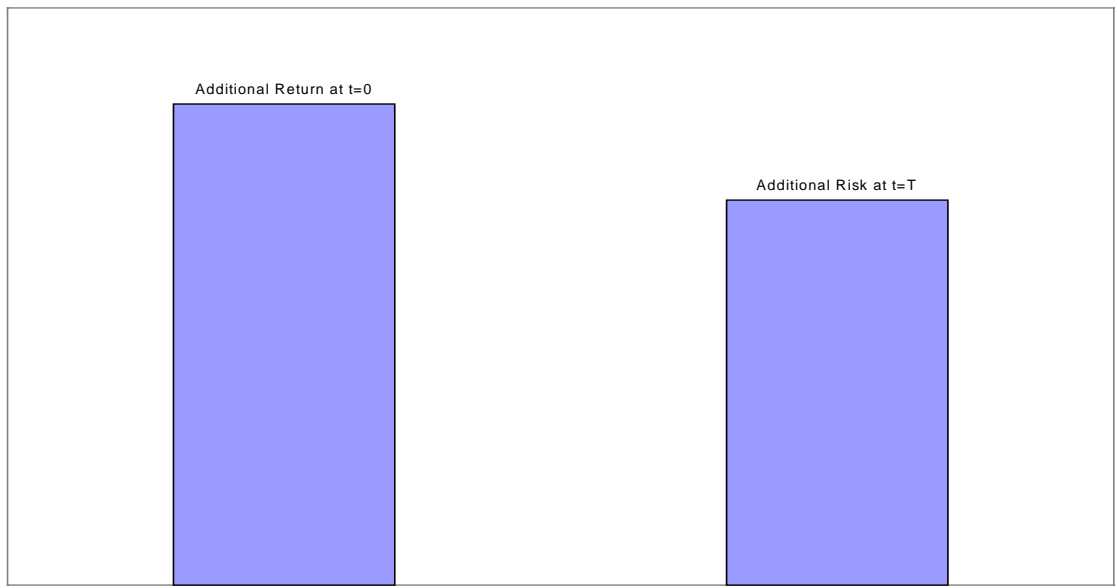

We need to find way to determine additional risk and additional return at the same time. In order to do this we need to find way to discount additional risk from time t= T to time t=0. Because the value of additional risk is money, we could suggest two discount procedures:

1. Additional Risk(t=0) = $e^{-rdt}$ Additional Risk(t = T)

(10)

2. Additional Risk ( t = 0) = $\frac{P_r}{I}$ Additional Risk(t = T)

The first statement from (10) corresponds to usual risk-neutral approach and the second to the self-agreeable discount procedure( if the value of $I$ (at time at present time costs Pr, then the value of additional risk at present time costs (Pr/$I$)*Additional Risk ( t = T) ). From this point on we will consider only the second approach. Finally, equation (1) could be rewritten as:

$$E - P_r = \alpha \frac{P_r^2}{I} e^{\mu^e \Delta t}(1 - e^{-\beta \sigma^e \sqrt{\Delta t}})$$

(11)

Equation (11) is the simplest square equation and because of this we could immediately write down its solution:

$$P_{r1,2} = -\frac{I}{2\alpha e^{\mu^e \Delta t}(1-e^{-\beta\sigma^e\sqrt{\Delta t}})} \pm \sqrt{\left(\frac{I}{2\alpha e^{\mu^e \Delta t}(1-e^{-\beta\sigma^e\sqrt{\Delta t}})}\right)^2 + \frac{EI}{\alpha e^{\mu^e \Delta t}(1-e^{-\beta\sigma^e\sqrt{\Delta t}})}}$$

(12)

Because Pr supposed to be positive, we will consider only (+) in (12).

We could simplify (12) for 2 different cases: big and small t:

1. small $\Delta t$

a>>b, where  (13)

$$a = \left(\frac{I}{2\alpha e^{\mu^e \Delta t}(1-e^{-\beta\sigma^e\sqrt{\Delta t}})}\right)^2, b = \frac{EI}{\alpha e^{\mu^e \Delta t}(1-e^{-\beta\sigma^e\sqrt{\Delta t}})}$$

(14)

2. and for big t- opposite inequality:

a << b  (15)

To simplify discussion for P let's write down dependence of from t (8, 9) in general form:

$$\sigma^e = \sigma(\Delta t)^\gamma$$

(16)

Where $\gamma$ =0, if $\sigma^e$ = constant, and $\gamma = 1$, if $\sigma^e \approx \Delta t$.

On the base of this equation we could get simplified expression for price of security Pr in the case of (13) and (14):

Small $\Delta t$:

$$P_r = E - \frac{E^2 \alpha e^{\mu^e \Delta t}}{I}(1 - e^{-\beta \sigma \Delta t^{\gamma + \frac{1}{2}}}) \tag{17}$$

or :

$$P_r = E - \frac{E^2 \alpha}{I} \beta \sigma \Delta t^{\gamma + \frac{1}{2}} \tag{18}$$

Now we could write equation for y (yield) using the yield definitions:

$$P_r = I e^{-y \Delta t} \tag{19}$$

and

$$E = I e^{-y_0 \Delta t}, \tag{20}$$

where $y_0$ – yield for risk – neutral valuation.

From (18), (19) and (20) we immediately could obtain:

$$y \Delta t = y_0 \Delta t + \alpha \beta \sigma \Delta t^{\gamma + \frac{1}{2}}, \tag{21}$$

where we left only two first powers of $\Delta t$ in (18). For forward rate we could obtain from (21) dependence on $\Delta t$, similar to formulas suggested in articles [4,5].

In particularly, we could obtain forward rate proportional to $\sqrt{\Delta t}$.

In the case of big $\Delta t$ (15) we could rewrite (12) as:

$$P_r = \sqrt{\frac{EI}{\alpha e^{\mu^e \Delta t}(1 - e^{-\beta \sigma^e \sqrt{\Delta t}})}} \tag{22}$$

or

$$e^{-y \Delta t} = e^{-\frac{\mu^e + y_0}{2} \Delta t}(1 - e^{-\beta \sigma^e \sqrt{\Delta t}})^{-\frac{1}{2}} \alpha^{-\frac{1}{2}} \tag{23}$$

Finally, $y \Delta t$:

$$y \Delta t = \frac{y_0 + \mu^e}{2} \Delta t - \frac{e^{-\beta \sigma^e \sqrt{\Delta t}}}{2} + \frac{\ln \alpha}{2} \tag{24}$$

( because we consider $e^{-\beta \sigma^e \sqrt{\Delta t}}$ as small term).

Equations (12), (18), (22), and derived from them equations (21) and (24) allow us to compare this model with existing financial data. We will use data from Federal Reserve Statistical Release [6] from 04/12/99 for U.S. treasury constant maturities.

The data (yields) in percents per annum are presented in table 1.

Table 1. Dependence of yields from maturity

| Time to maturity | Yields in percent per annum |
|---|---|
| 3m | 4.32 |
| 6m | 4.5 |
| 1y | 4.66 |
| 2y | 4.93 |
| 3y | 4.95 |
| 5y | 4.98 |
| 7y | 5.17 |
| 10y | 5.06 |
| 20y | 5.72 |
| 30y | 5.45 |

## 4. Possible strong hypothesis about relationship between observed and risk-neutral interest rates and comparison with existing financial data.

In order to compare obtained result with financial data we need model for calculating y (t) and another parameters from (12). One of the most extreme hypothesis could suggest that

$$y_0(t) = constant = y_0 \qquad (25)$$

Because of it we will be able to try to compare obtained time dependencies (21) and (24) with yields, obtaining from bond prices.

To do this, we compare data from table (1) and suggested dependencies (21) – for small times and (24) – for big times.

Let's first of all discuss comparison between suggested financial data and formula (21): If we choose $\gamma$ in (21) equal to 1, than formula (21) could be rewritten as:

$$y = y_0 + \alpha\beta\sigma\sqrt{\Delta t} \qquad (26)$$

In order to compare suggested function with financial data we will build chart of dependence y from $\sqrt{\Delta t}$ (Fig.2).

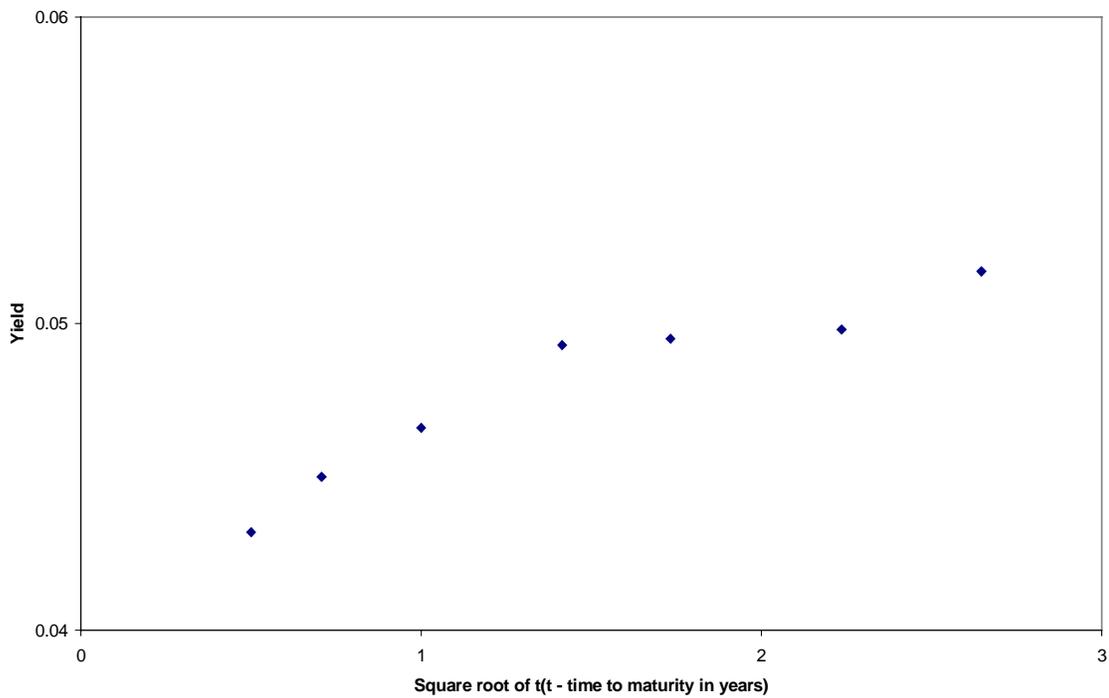

**Fig.2 Dependence of yield from time to maturity**

It's clear from this dependence that there is a reasonably good agreement between suggested formula (26) and actual dependence y(t). From this graphic (Fig.2) we could determine coefficients in formula (26): $y_0$ and $\alpha\beta\sigma$ :

$$y_0 = 0.004, \text{ and } \alpha\beta\sigma = 0.66*10^{-2} \qquad (27)$$

Now we could try to compare (24) with data from table 1. In order to do it we will introduce some additional assumptions: $\mu^e = const$, and $\mu^e = y_0$ - there is only one rate of return and this rate is equal to $y_0$.

Now, assuming that $e^{-\beta\sigma^e\sqrt{\Delta t}}$ is small enough for last 2 existing values (20 years and 30 years) we could simplify (24) and write it as:

$$y\Delta t = y_0\Delta t + \frac{\ln \alpha}{2} \qquad (28)$$

Using values of y for these two times to maturity (20 and 30 years), and knowing that $y_0 = 0.004$, we could determine (Fig.3):

$$\alpha = 1.18 \qquad (29)$$

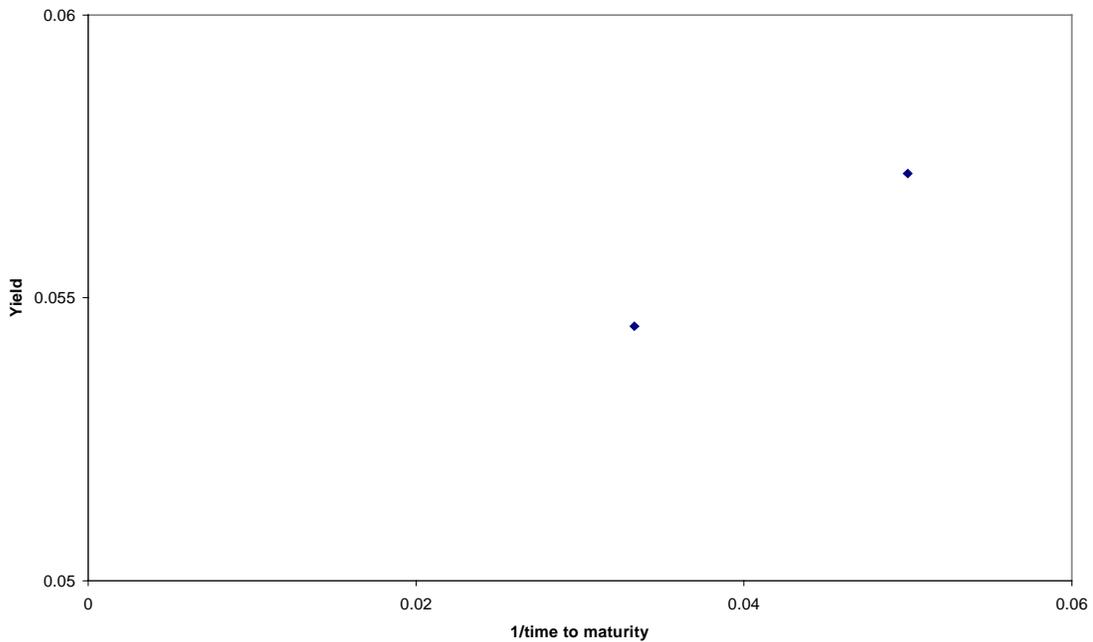

Fig.3 Simple parameter determination

Then, from (24) we could receive the following formula:

$$\ln(-y\Delta t + y_0\Delta t + \frac{\ln \alpha}{2}) + \ln 2 = -\beta\sigma^e\sqrt{\Delta t} \qquad (30)$$

Now, because we know $\alpha$ (1.18) and $\alpha\beta\sigma$ (0.66*10$^{-2}$), it's possible to compare coefficients $\beta\sigma$ from (28) with coefficients $\beta\sigma$, obtained from (27) and (29) ($\beta\sigma^e = \beta\sigma\Delta t$):

$$\beta\sigma^e = \frac{\alpha\beta\sigma^e}{\alpha} \qquad (31)$$

Here, also we present data, calculated using equation (12) for different parameters $\alpha$ ($\alpha = 1.18$ and $\alpha = 2.05$ - tables 2 and figures 4, 5 correspondingly):

It's clear from these data that difference between existing price and price, calculated on the base of suggested model is no more than 5%. Even additional precision in parameters' estimation could decrease this difference. In addition, for better correspondence between observed data and suggested model, it's possible to relax the following assumptions: $\mu^e = const, \mu^e = y_0, y_0 = const, \gamma = 1$.

Table2. Prices of treasury securities and calculated prices ( for $\alpha = 1.18$ and $\alpha = 2.05$).

| Time to maturity (years) | Price of treasury security | Calculated price of treasury security 1 | Calculated price of treasury security 2 |
|---|---|---|---|
| 0.25 | 0.9893 | 0.9892 | 0.9892 |
| 0.5 | 0.9778 | 0.9779 | 0.9779 |
| 1 | 0.9545 | 0.9545 | 0.9545 |
| 2 | 0.9061 | 0.9066 | 0.9066 |
| 3 | 0.8620 | 0.8588 | 0.8586 |
| 5 | 0.7796 | 0.7673 | 0.7667 |
| 7 | 0.6964 | 0.6839 | 0.6826 |
| 10 | 0.6029 | 0.5757 | 0.5731 |
| 20 | 0.3185 | 0.3340 | 0.3272 |
| 30 | 0.1950 | 0.2036 | 0.1949 |

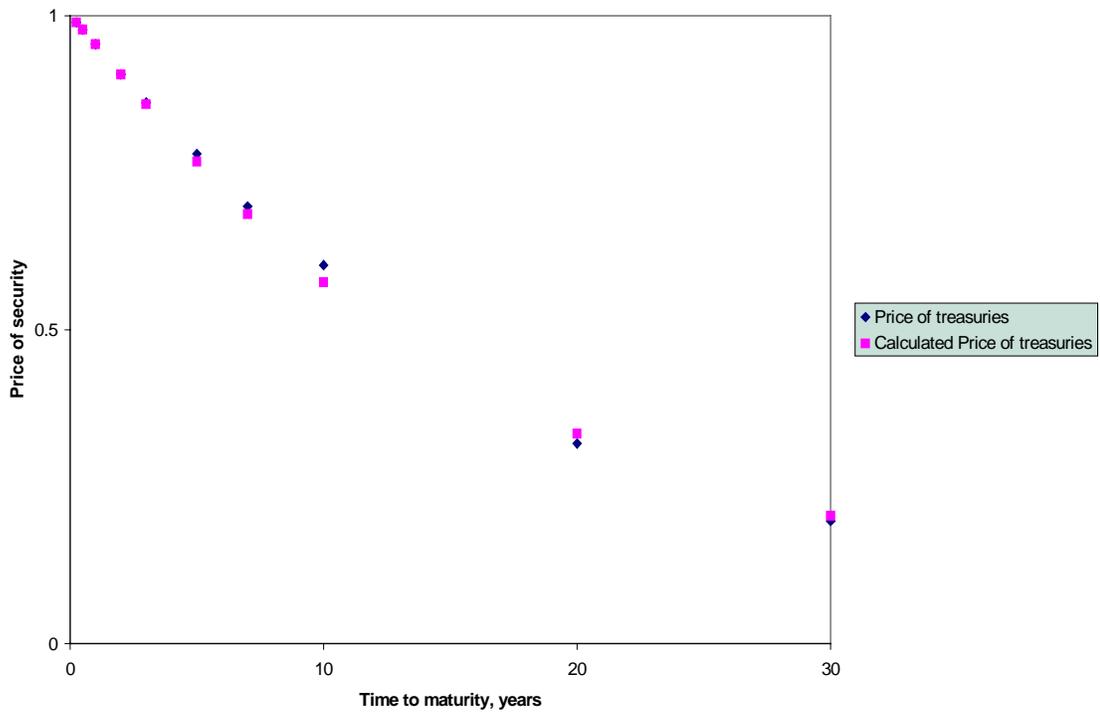

Fig.4 Dependence of treasuries prices from time

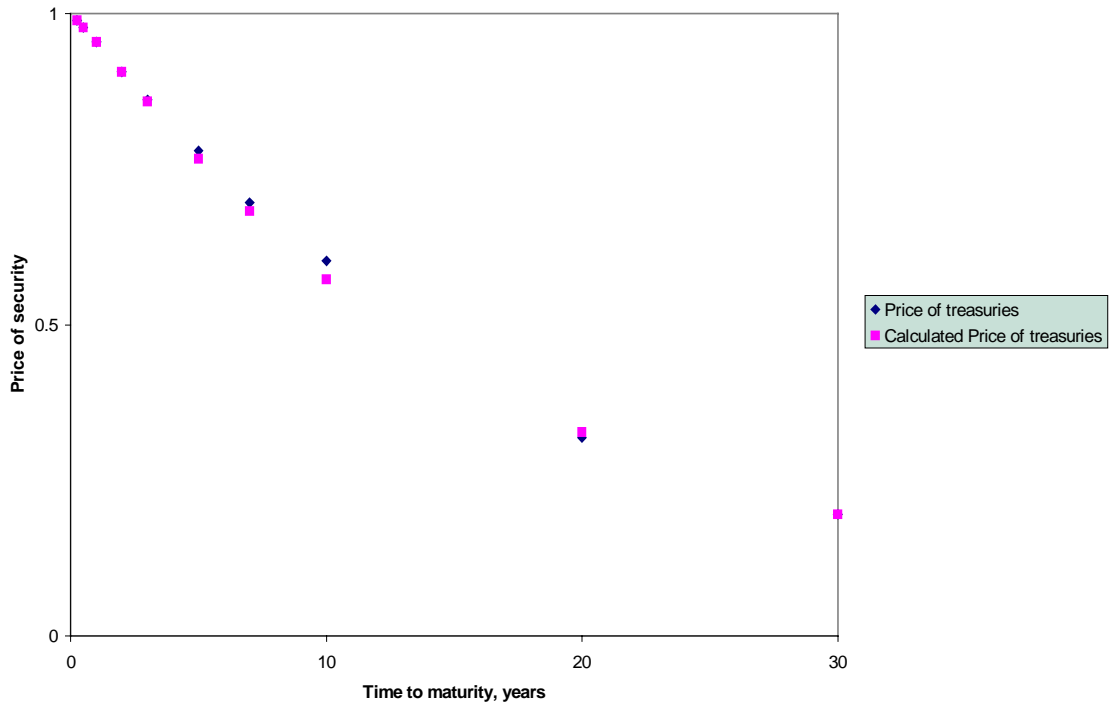

**Fig.5 Dependence of treasuries prices from time**

Calculated price of treasury security 1 corresponds to $\alpha = 1.18$ and calculated price of treasury security 2 corresponds to $\alpha = 2.05$.

## 5. Portfolio theory, risk - neutral model and interest - rate models.

Here we will show relationship between existing financial models and suggested way for mathematical formulation of liquidity preference theory.
Let's point out again, that choosing of VaR as risk measure, allow us to evaluate risk for different maturities.

Additional return could be calculated as difference between risk - neutral price ( price, calculated on the base of risk - neutral model and market price). To connect return and risk for different maturities we will use relationship between risk and return, similar to portfolio theory. Finally, we will express suggested formalism, using graphical view:

$$\textit{Formula, connected with portfolio theory} =$$
$$\text{(Formula, connected with risk - neutral valuation)} /$$
$$\text{(Formula, connected with VaR)};$$

### 6. Conclusion.

New mathematical formulation of liquidity preference theory is suggested. On the base of comparison between suggested model and real prices paradoxical conclusion could be derived. All yield curve could be described only on the base of liquidity preference theory without any consideration of risk-neutral models.